\documentclass[%
prc,%
10pt,%
final,%
notitlepage,%
oneside,%
twocolumn,%
nobibnotes,%
nofootinbib,
superscriptaddress,%
floatfix,%
floatfix,%
showkeys,%
showpacs]%
{revtex4}
\usepackage{color}
\usepackage{amsfonts}
\usepackage{amsbsy}
\usepackage{mathrsfs}
\usepackage{graphicx}
\def\lsim{\mathrel{\rlap{
\lower4pt\hbox{\hskip-3pt$\sim$}}
    \raise1pt\hbox{$<$}}}     
\def\gsim{\mathrel{\rlap{
\lower4pt\hbox{\hskip-3pt$\sim$}}
    \raise1pt\hbox{$>$}}}     

\begin{document}
\title{	
Global $\Lambda$ polarization in heavy-ion collisions at high baryon density
} 
\author{Yu. B. Ivanov}\thanks{e-mail: yivanov@theor.jinr.ru}
\affiliation{Bogoliubov Laboratory for Theoretical Physics, 
Joint Institute for Nuclear Research, Dubna 141980, Russia}
\affiliation{National Research Centre "Kurchatov Institute",  Moscow 123182, Russia} 
\begin{abstract}
Based on the model of three-fluid dynamics (3FD), the global $\Lambda$  polarization ($P_\Lambda$) is calculated
in Au+Au collisions at 3 $\leq\sqrt{s_{NN}}\leq$ 9 GeV, in which   
high baryon density is achieved. 
Various contributions to $P_\Lambda$ are considered: those from  
the thermal vorticity, meson field, thermal shear and spin-Hall effect. 
Feed-down from higher-lying resonances is also taken into account.
The results are compared with available data. 
Special attention is payed to the collision energies of 
$\sqrt{s_{NN}}=$ 3, 3.2, 3.5, 3.9, and 4.5 GeV, 
for which a thorough scan of the energy, 
rapidity, and centrality dependence of $P_\Lambda$ is performed.
The results for 3 GeV reasonably well reproduce the corresponding STAR data. 
While the results at $\sqrt{s_{NN}}=$ 3.2, 3.5, 3.9, and 4.5 GeV
can be considered as predictions for results of measurements 
within the STAR fixed-target (STAR-FXT)  program
that are expected in the nearest future.
It is predicted that 
a broad maximum of $P_\Lambda$ is reached at $\sqrt{s_{NN}}\approx$ 3--3.9 GeV,  
exact position of which depends on the centrality and 
width of the midrapidity range of observation. 
Impact of the meson-field, thermal-shear  and spin-Hall-effect contributions 
to  $P_\Lambda$ is also studied. 
\end{abstract}
\maketitle

\section{Introduction}
\label{Introduction}

Non-central heavy-ion collisions are characterized by huge angular momenta of the order of 10$^3$--10$^4 \hbar$. 
Although a part of the angular momentum is carried away
by spectators, a large fraction of it remains in the participant region of the colliding nuclei 
and induces collective vortical motion of the participant matter. 
This vortex motion is partially transformed into a preferential orientation of spins of the participant 
particles through spin-orbit coupling, which can be observed as a polarization (for fermions) or alignment (for vector mesons) of emitted particles. Observation of this polarization (alignment) gives us access to information about the vortex motion during heavy-ion collisions, provided the mechanism of partial transformation of the collective vortex motion into the spin polarization (alignment) is understood. 
The current state of theoretical and experimental research in this area is discussed in recent reviews 
\cite{Niida:2024ntm,Becattini:2024uha}.

Measurements of the global polarization of the $\Lambda$ and  $\bar{\Lambda}$ hyperons
by the STAR Collaboration at collision energies 7.7 $\leq \sqrt{s_{NN}} \leq$ 39 GeV
and at $\sqrt{s_{NN}}=$ 200 GeV  showed statistically significant  result \cite{STAR:2017ckg,Adam:2018ivw}. 
These measurements demonstrated a rising global polarization  with decreasing $\sqrt{s_{NN}}$.
A simple extrapolation of this trend would suggest that the global polarization
continues to rise as $\sqrt{s_{NN}}$ decreases. However, the global polarization
should vanish at $\sqrt{s_{NN}}=2m_N$ ($m_N$ is the nucleon mass) due to zero 
angular momentum of the system at this energy.
Therefore, a peak of the global polarization should occur in the region $2m_N <\sqrt{s_{NN}}\leq$ 7.7 GeV, 
where  high baryon density is reached.

Various predictions for the global $\Lambda$ polarization ($P_\Lambda$) at $\sqrt{s_{NN}}<$ 7.7 GeV were done 
\cite{Deng:2020ygd,Ivanov:2019ern,Ivanov:2019wzg,Ivanov:2020udj,Guo:2021udq,Voronyuk:2023vyu,Ayala:2021xrn}. 
The maximum $P_\Lambda$ has
been predicted around 7.7 GeV  in Ref. \cite{Guo:2021udq} while the averaged thermal vorticity, 
i.e. the driving force of the polarization, was reported to reach a peak at around 3--4 GeV \cite{Deng:2020ygd,Guo:2021udq}. 
Calculations of Refs. \cite{Ivanov:2019ern,Ivanov:2019wzg,Ivanov:2020udj,Ayala:2021xrn} predicted 
monotonous rise (with decreasing $\sqrt{s_{NN}}$) to $\approx$3 GeV. 
The monotonous increase  
to $\approx$3--4 GeV was also predicted in Ref. \cite{Voronyuk:2023vyu}. 
However, this is a much more moderate increase that essentially underestimates the STAR data at 3 GeV \cite{STAR:2021beb}. 
Ref. \cite{Ayala:2021xrn} reported almost constant $P_\Lambda$ from 7.7 to $\approx$3 GeV and then rapid rise to 
maximum at $\sqrt{s_{NN}}=$  3 GeV followed by rapid fall to zero.

The STAR \cite{STAR:2021beb} and HADES \cite{HADES:2022enx} data on $P_\Lambda$ 
at $\sqrt{s_{NN}}=$ 3 GeV and $\sqrt{s_{NN}}=$ 2.4 GeV, respectively, showed that 
predictions of Refs. \cite{Ivanov:2020udj,Ayala:2021xrn} capture the experimental trend.
These data demonstrated that increase 
of $P_\Lambda$  with decreasing $\sqrt{s_{NN}}$ continues below 7.7 GeV 
and  indicated that the peak in $P_\Lambda$ is reached definitely below the energy of 7.7 GeV.
However, the questions 
remained: is it a gradual increase from 7.7 to 3 GeV or with a peak in between, or is it a slow increase to 3 GeV and then steep fall \cite{Ayala:2021xrn}?
As aforementioned, these different scenarios reflect different pattens of the collective vortical motion in 
the produced nuclear matter. These different pattens in-particular lead to differences in rapidity dependence of $P_\Lambda$
\cite{Guo:2021udq,Ivanov:2022ble}. 

In Ref. \cite{Ivanov:2022ble}, the calculations in the energy range 2.4 $<\sqrt{s_{NN}}<$ 7.7 GeV have been 
already performed but it was not a thorough scan of this energy range. The main purpose of Ref. \cite{Ivanov:2022ble}
was to study  meson-field induced contribution to the global polarization proposed in 
Ref.  \cite{Csernai:2018yok}, see also \cite{Xie:2019wxz}, 
and its application to description of data at $\sqrt{s_{NN}}=$ 2.4, 3 and 7.2 GeV
\cite{STAR:2021beb,HADES:2022enx,Okubo:2021dbt}. 
%
Recently it has been found that there are additional contributions to the 
spin polarization at local equilibrium \cite{Becattini:2024uha,Becattini:2021suc,Liu:2021uhn,Liu:2020dxg,Becattini:2021iol}.
These are a so-called thermal-shear  and spin-Hall-effect contributions,
which await evaluation in the STAR-FXT energy range.

In the present paper,  the global $\Lambda$  polarization is calculated
in Au+Au collisions at 3 $\leq\sqrt{s_{NN}}\leq$ 9 GeV. 
Special attention is payed to the collision energies of 
$\sqrt{s_{NN}}=$ 3, 3.2, 3.5, 3.9, and 4.5 GeV, 
for which a thorough scan of the energy, 
rapidity, and centrality dependence of $P_\Lambda$ is performed.
Effects of the thermal-shear  and spin-Hall-effect contributions \cite{Liu:2021uhn,Liu:2020dxg}
to  $P_\Lambda$ are also studied. 
These simulations closely follow the procedure described in Ref. \cite{Ivanov:2022ble}. 
The obtained results at energies of 3.2--4.5 GeV can be considered as a prediction for the STAR-FXT
experimental data that are expected in the nearest future.  
These results are also relevant to forthcoming experiments 
at Nuclotron-based Ion Collider fAcility (NICA) in Dubna (BM@N and MPD experiments).

\section{Global polarization in 3FD model}
\label{polarization in 3FD model}

The scheme of the $P_\Lambda$  calculation is precisely the same as that described in Ref. 
\cite{Ivanov:2022ble}. In Ref. \cite{Ivanov:2022ble} two contributions to 
the global $\Lambda$ polarization were considered. The first one is the conventional contribution 
associated with the thermal vorticity 
   \begin{eqnarray}
   \label{therm.vort.}
   \varpi_{\mu\nu} = \frac{1}{2}
   (\partial_{\nu} \beta_{\mu} - \partial_{\mu} \beta_{\nu}). 
   \end{eqnarray}
Here $\beta_{\mu}=u_{\nu}/T$, 
$u_{\mu}$ is collective local four-velocity of the matter,  and
$T$ is local temperature.  Here $u_{\mu}$ and $T$ are atributed to the unified fluid
in terms of the 3FD model 
because the system is equilibrated at the freeze-out stage. 
It is instructive to divide thermal vorticity into kinematic vorticity, $\omega_{\mu\nu}$,
and vorticity due to $1/T$ derivatives, $\omega_{\mu\nu}^{(T)}$
   \begin{eqnarray}
   \label{therm.vort.1}
   \varpi_{\mu\nu} =  \omega_{\mu\nu} + \omega_{\mu\nu}^{(T)},
   \end{eqnarray}
where
   \begin{eqnarray}
   \label{therm.vort.kin}
\omega_{\mu\nu} = \frac{1}{2T} (\partial_{\nu} u_{\mu} - \partial_{\mu} u_{\nu}),
   \end{eqnarray}
   \begin{eqnarray}
   \label{therm.vort.T}
\omega_{\mu\nu}^{(T)} = \frac{1}{2} (u_{\mu}\partial_{\nu} - u_{\nu}\partial_{\mu})\frac{1}{T}.
   \end{eqnarray}

The corresponding mean spin vector of $\Lambda$  particles with four-momentum $p$, 
produced around point $x$ on freeze-out hypersurface is
   \begin{eqnarray}
\label{xp-pol}
 S^{\varpi}_\mu(x,p)
 =\frac{1}{8m_\Lambda} [1-f_\Lambda (x,p)] \: \epsilon_{\mu\nu\rho\sigma} p^\sigma 
  \varpi^{\rho\nu}(x) 
   \end{eqnarray}
where $f_\Lambda(x,p)$ is the Fermi-Dirac distribution function and $m_\Lambda$ is mass of the 
$\Lambda$ hyperon. Further on, the Fermi factor $[1-f_\Lambda (x,p)]$ is omitted 
because the $\Lambda$'s are produced in high-temperature regions, where Boltzmann
statistics dominates.

The second, less conventional contribution, considered in Ref. \cite{Ivanov:2022ble}, 
is related to the meson-field contribution \cite{Csernai:2018yok,Xie:2019wxz}. 
The corresponding mean spin vector is 
\begin{eqnarray}
\label{pixpgen2}
   S^{V}_\mu(x,p) =  \frac{1}{8m_\Lambda} \epsilon_{\mu\nu\rho\sigma} p^\sigma 
  \left( \beta_\Lambda \frac{g_{V\Lambda}}{m_\Lambda T}\right) V^{\rho\nu} ,
\end{eqnarray}
where the Fermi factor 
$[1-f_\Lambda(x,p)]$ is again omitted,   
$\beta_\Lambda =1$ [note that $\beta_{\bar{\Lambda}}=-1$], 
$g_{V\Lambda}$ is coupling constant of the vector meson with $\Lambda$ hyperon, 
\begin{eqnarray}
V_{\mu\nu} = \partial_\mu V_\nu - \partial_\nu  V_\mu 
\label{e1}
\end{eqnarray}
is the vector-field strength tensor, and the $V$ field itself is defined as 
\begin{eqnarray}
 V^\nu =  \frac{\bar{g}_{V}}{m^2_{V}} J^\nu_B 
\label{sigma-omega-sol}
\end{eqnarray}
through the baryon current 
$J^\nu_B=n_B u^\nu$, $n_B$ is the baryon density, $m_{V}$ is the vector-meson mass, and 
$\bar{g}_{V}$ is the mean coupling constant of the vector meson with various baryons. 
The vector-meson-field contribution (\ref{pixpgen2}) is similar to that of the electromagnetic field interacting 
with magnetic moment of the hyperon \cite{Becattini:2016gvu}.

The meson-field contribution to the mean spin vector (\ref{pixpgen2}) 
was derived in Ref. \cite{Csernai:2018yok,Xie:2019wxz} based on the 
relativistic mean-field (RMF) model \cite{Serot-Walecka}, where the vector meson is associated with 
the $\omega$ meson. Therefore $m_{V}=m_\omega=$ 783 MeV.  
Values of coupling constants differ in different realizations of the  
RMF model. Following Refs. \cite{Csernai:2018yok,Xie:2019wxz}, 
just one of the possible parametrizations is used: 
$\bar{g}_{V}=g_{VN}=$ 8.646 and $g_{V\Lambda}=0.67 g_{VN}$ \cite{Cohen:1991qi}. 
The mean coupling constant is associated with the nucleon one because nucleons 
dominate in the system at low energies considered here.

The polarization of the $\Lambda$ hyperon is measured in
its rest frame, therefore the $\Lambda$ polarization is 
   \begin{eqnarray}
   \label{P_L-rest}
  \Pi^\mu_{\Lambda}(x,p) =  S^{*\mu}(x,p)/S _{\Lambda} 
   \end{eqnarray}
where $S _{\Lambda}=$ 1/2 is the spin of the $\Lambda$ hyperon and
$S^{*\mu}_{\Lambda}$ is  mean $\Lambda$-spin vector its rest frame. 
The zeroth component  $S^{0}_{\Lambda}$  identically vanishes 
 in the $\Lambda$ rest frame. The $y$ component of $\Pi^\mu_{\Lambda}$
is of prime interest
because the global polarization is directed orthogonally to the reaction plane ($xz$).

Particles are produced across entire
freeze-out hypersurface. Therefore, to calculate the global
polarization ($P_{\Lambda}$), the above
expression should be averaged over the freeze-out hypersurface $\Sigma$
and particle momenta 
   \begin{eqnarray}
\label{polint}
 P_{\Lambda}
 = \frac{\int (d^3 p/p^0) \int_\Sigma d \Sigma_\lambda p^\lambda
f_\Lambda   \Pi^y_{\Lambda}}
 {\int (d^3 p/p^0) \int_\Sigma d\Sigma_\lambda p^\lambda \, f_\Lambda}. 
   \end{eqnarray}
In principle, the integration over momenta should run over the range of the 
experimental acceptance. However, here the integration runs over all momenta, while  
the experimental rapidity acceptance is approximately taken into account in terms of a so-called 
hydrodynamical rapidity 
   \begin{eqnarray}
   \label{y}
y_h = \frac{1}{2} \ln \frac{u^0+u^3}{u^0-u^3} , 
   \end{eqnarray}
based on hydrodynamical 4-velocity $u^\mu$. It means that the  $d\Sigma_\lambda p^\lambda$ integration runs 
only over part of the freeze-out hypersurface, where condition $|y_h|<y_{\rm{acceptance}}$ is met. 
This part the hypersurface is denoted as $\Sigma_{\Delta y}$.  
Similarly to previous 3FD calculations of the $\Lambda$ polarization 
\cite{Ivanov:2019ern,Ivanov:2019wzg,Ivanov:2020udj,Ivanov:2022ble},
a simplified version of the freeze-out is used, i.e. an isochronous one. 
The freeze-out instant is taken to be equal to that, when the average energy density 
in the central region (i.e. slab $|z|\leq$ 4 fm) drops down to 
the value of the average freeze-out energy density in the same central region
obtained in conventional 3FD simulations with differential, i.e.
cell-by-cell, freeze-out \cite{Russkikh:2006aa,Ivanov:2008zi}.

Performing integration over $dp$ as it is described in Ref. \cite{Ivanov:2022ble} in detail, 
one arrives to the following expressions for thermal-vorticity ($P_{\Lambda}^\varpi$) 
and meson-field ($P_{\Lambda}^V$) contributions
to the global $\Lambda$ polarization: 
   \begin{eqnarray}
\label{polint-fin}
 P_{\Lambda}^\varpi
 = \frac{1}{6} \frac{\int_{\Sigma_{\Delta y}} d^3 x 
 \left(\rho_{\Lambda} +  2 T^{00}_{\Lambda}/m_{\Lambda} \right) 
\varpi_{zx}}%
{\int_{\Sigma_{\Delta y}}  d^3 x \, \rho_{\Lambda}}, 
   \end{eqnarray}
   \begin{eqnarray}
\label{polint-V-fin}
 P_{\Lambda}^V
 = 
\frac{\beta_\Lambda g_{V\Lambda}}{6m_\Lambda T}
\frac{\int_{\Sigma_{\Delta y}} d^3 x 
 \left(\rho_{\Lambda} +  2 T^{00}_{\Lambda}/m_{\Lambda} \right) 
V_{zx}}%
{\int_{\Sigma_{\Delta y}}  d^3 x \, \rho_{\Lambda}}, 
   \end{eqnarray}
where $\rho_{\Lambda}$ is the $\Lambda$ density in the frame of calculation and 
$T^{00}_{\Lambda}$ is the $00$ component of the partial energy-momentum tensor related to 
the $\Lambda$ contribution 
   \begin{eqnarray}
   \label{T00Heq}
T^{\tau \lambda}_{\Lambda} = (\varepsilon_{\Lambda} + p_{\Lambda})u^\tau u^\lambda - g^{\tau \lambda} p_{\Lambda}
   \end{eqnarray}
with $\varepsilon_{\Lambda}$ and $p_{\Lambda}$ being the corresponding partial 
energy density and pressure, respectively. 
$\rho_{\Lambda}$, $\varepsilon_{\Lambda}$ and $p_{\Lambda}$ are determined by ideal-gas 
relations in terms of temperature, baryon and strange chemical potentials because 
the system is described by the ideal-gas EoS at the freeze-out stage.

These are the final expressions with which simulations are performed. 
The feed-down from higher-lying resonances is also taken into account, as it is 
described in Ref. \cite{Ivanov:2022ble} in detail.

\section{Local Equilibrium}
\label{Local}

In addition to the thermal vorticity    (\ref{polint-fin}) and 
meson-field (\ref{polint-V-fin}) terms, 
there are other contributions to the spin polarization vector in the local equilibrium 
\cite{Becattini:2024uha,Becattini:2021suc,Liu:2021uhn,Liu:2020dxg,Becattini:2021iol}. 
These are so-called shear-induced polarization (SIP) 
\begin{eqnarray}
   \label{SIP}
S^{\rm{SIP}}_\mu(x,p)
 =\frac{1}{4m_\Lambda}
\frac{p^{\nu} u^{\beta}p_{\rho}}{(u\cdot p)} 
\epsilon_{\mu\nu\alpha\beta}\xi^{\rho\alpha}
\end{eqnarray}
and polarization induced by the spin-Hall effect (SHE)
\begin{eqnarray}
   \label{SHE}
S^{\rm{SHE}}_\mu(x,p)
 =\frac{1}{4m_\Lambda}
\frac{p^{\alpha}u^{\beta}}{(u\cdot p)}\epsilon_{\mu\nu\alpha\beta}\partial^{\nu}\frac{\mu}{T},
\end{eqnarray}
where  
\begin{eqnarray}
   \label{xi}
  \xi_{\mu\nu} = \frac{1}{2} \left( \partial_\mu \beta_\nu + \partial_\nu \beta_\mu \right) 
\end{eqnarray}
is the {\em thermal shear} tensor, defined as the symmetric derivative of the four-temperature, 
and $\mu$ is the chemical potential. 
These new terms, SIP and SHE, are related to the motion of 
the particle in anisotropic fluid. They become identically zero in the global equilibrium. 
To be precise, the above expressions are presented in the 
explicitly covariant form of Refs. \cite{Liu:2021uhn,Liu:2020dxg}, which 
results from quantum kinetic theory.

As mentioned above, the polarization of the $\Lambda$ hyperon is measured in
its rest frame, see Eq. (\ref{P_L-rest}). 
Therefore, the mean spin vector in the $\Lambda$ rest frame, $S^{*\mu}$, is needed. 
In the  $\Lambda$ rest frame the zeroth component  $S^{*0}$ is identically zero 
and the spatial component becomes 
   \begin{eqnarray}
\label{S-rest}
 {\bf S}^*(x,p)
 = {\bf S} -  \frac{S^0}{p^0+m_{\Lambda}} {\bf p} 
\stackrel{def}{=}
 {\bf S} - \Delta{\bf S}
   \end{eqnarray}
where  relation ${\bf p} \cdot {\bf S}=p^0 S^0$ was used
to modify the conventional expression for $S^{*\mu}$.
Here, $p = (p^0,{\bf p})$ is the four-momentum of the emitted  $\Lambda$ hyperon  
and $ S_\mu = S^\varpi_\mu +  S^V_\mu +  S^{\rm{SIP}}_\mu + S^{\rm{SHE}}_\mu$ 
is the mean spin vector in the frame of calculation, i.e. the center-of-mass frame of colliding nuclei. 
$\Delta{\bf S}$ is the boost term that originates from the boost of 
the mean spin vector to the $\Lambda$ rest frame.

As aforementioned, the the mean spin vector in the $\Lambda$ rest frame, $S^{*\mu}/S_\Lambda$, 
should be averaged over the freeze-out hypersurface and particle momenta in order to calculate the global polarization, see Eq. (\ref{polint}). 
The thermal-vorticity and meson-field contributions are presented in 
Eqs. (\ref{polint-fin}) and (\ref{polint-V-fin}). 
The momentum integration of ${\bf S}^{\rm{SIP}}$ and ${\bf S}^{\rm{SHE}}$
gives zero \cite{Ivanov:2022geb}
\begin{eqnarray}
\label{SIP=0}
\int \frac{d^3 p}{p^0} \int d\Sigma^{\lambda}p_{\lambda}f_\Lambda
S^{\rm{SIP}}_{\mu} 
& = & 
0,
\\
\label{SHE=0}
\int \frac{d^3 p}{p^0} \int d\Sigma^{\lambda}p_{\lambda}f_\Lambda
S^{\rm{SHE}}_{\mu} 
& = & 
0.
\end{eqnarray}
A nonzero result may be produced by a finite momentum acceptance.

The corresponding boost corrections $\Delta{\bf S}^{\rm{SIP}}$ and $\Delta{\bf S}^{\rm{SHE}}$, 
i.e. the partial SIP and SHE contributions to the above boost term $\Delta{\bf S}$, 
give nonzero results but they contain additional smallness $|{\bf p}|/(p^0+m_{\Lambda})$. 
This is really small parameter in midrapidity, of the order of $\sim T/m_{\Lambda}$. 
At $y\approx 1$ it is $\approx 0.3$ and tends to 1 at $y\gg 1$. Here, use is made of the approximation 
$1/(p^0+m_{\Lambda})\approx 1/[m_{\Lambda}(u^0+1)]$. 
The accuracy of this approximation is $\sim T/m_{\Lambda}$. 
This replacement allows one to immediately perform the momentum integration and arrive at the 
$\Delta{\bf S}^{\rm{SIP}}$ and $\Delta{\bf S}^{\rm{SHE}}$ contributions to $P_\Lambda$:
\begin{eqnarray}
\label{P_SIP}
P_\Lambda^{\rm{SIP}}  
&\simeq& 
-\frac{1}{2m_{\Lambda}}
\int_{\Sigma_{\Delta y}} d^3 x  \frac{u^0p_\Lambda}{(u^0+1)}  \Xi_{zx}
/ \int d^3 x \rho_\Lambda,
\\
\label{P_SHE}
P_\Lambda^{\rm{SHE}}  
&\simeq& 
\frac{1}{m^2_{\Lambda}}
\int_{\Sigma_{\Delta y}} d^3 x   \frac{u^0p_\Lambda}{(u^0+1)}  \zeta_{zx}
/ \int d^3 x \rho_\Lambda,
\end{eqnarray}
where 
\begin{eqnarray}
\label{Xi}
\Xi_{\mu\nu} =    
u_\mu u^\lambda \xi_{\nu\lambda} - u_\nu u^\lambda \xi_{\mu\lambda},
\end{eqnarray}
\begin{eqnarray}
\label{zeta_}
\zeta_{\mu\nu} = \frac{1}{2} 
\left(u_\mu  \partial_{\nu}\frac{\mu}{T} - u_\nu  \partial_{\mu}\frac{\mu}{T}\right),
\end{eqnarray}
and $p_{\Lambda}$ is the partial pressure of $\Lambda$ hyperons, see Eq. (\ref{T00Heq}) 
The chemical potential is 
$\mu=B\mu_B+S\mu_S$, where $\mu_B$ and $\mu_S$ are baryon and strangeness chemical potentials, respectively, 
$B$ and $S$ are baryon and strangeness charges of the considered baryon. For $\Lambda$ hyperon 
$\mu=\mu_B-\mu_S$. 
Here $d^3 x$ integration runs over cells, where condition $|y_h|<y_{\rm{acceptance}}$ is met. 
This part the hypersurface is denoted as $\Sigma_{\Delta y}$, similarly to that in Eqs. 
(\ref{polint-fin}) and (\ref{polint-V-fin}). 

The thermal-shear vorticity (\ref{Xi}) can be subdevided into kinematic part, $\widetilde{\Xi}_{\mu\nu}$,
and the part due to $T$ derivatives, similarly to that in Eq. (\ref{therm.vort.1}),  
\begin{eqnarray}
\label{Xi1}
\Xi_{\mu\nu} =  \widetilde\Xi_{\mu\nu} + \omega_{\mu\nu}^{(T)},  
\end{eqnarray}
\begin{eqnarray}
\label{Xi-kin}
\widetilde{\Xi}_{\mu\nu} = 
\frac{1}{2T} 
\left[
u_\mu (u^\lambda \partial_\lambda) u_\nu   
-
u_\nu   (u^\lambda  \partial_\lambda) u_\mu  
\right]. 
\end{eqnarray}
It is seen that the part of $\Xi_{\mu\nu}$ due to $T$ derivatives
is precisely the same as that in the thermal vorticity (\ref{therm.vort.T}). 
It does not mean that their contributions to the global polarization are equal
because they are integrated with different weights over the freeze-out hypersurface.

Comparing expressions (\ref{P_SIP}) and (\ref{P_SHE}) with that for the thermal vorticity (\ref{polint-fin}), 
one sees that the SIP and SHE contributions indeed contain additinal smallness 
$u^0p_\Lambda/\rho_\Lambda   \sim T/m_{\Lambda}$ compared to $P_{\Lambda}^\varpi$, 
as it was mentioned above.

\section{Results}
\label{Results}

The global $\Lambda$ polarization in Au+Au collisions 
is considered in the energy range 3 $\leq\sqrt{s_{NN}}\leq$ 9 GeV, 
which is relevant to the STAR-FXT experiment and upcoming BM@N and MPD experiments at NICA. 
This energy range has been already studied in Ref. \cite{Ivanov:2022ble}.
Here a more systematic study is presented. Special attention is payed to the collision energies of 
$\sqrt{s_{NN}}=$ 3.2, 3.5, 3.9, and 4.5 GeV. The corresponding results can be considered as predictions 
for the STAR-FXT experimental data that are expected in the nearest future.

The simulations closely follow the procedure described in Ref. \cite{Ivanov:2022ble}. 
They are performed within the 3FD model \cite{3FD,Ivanov:2013wha} 
combined with thermodynamic approach to the particle polarization 
\cite{Becattini:2013fla,Becattini:2016gvu,Fang:2016vpj} 
supplemented by the meson-field induced contribution  \cite{Csernai:2018yok}, as well as  
the SIP and SHE contributions \cite{Liu:2021uhn,Liu:2020dxg} that are present 
in the local equilibrium. 
All results presented below are calculated taking into account feed-down from higher-lying resonances, 
based on prescription of Ref. \cite{Ivanov:2022ble}. 
The feed-down reduces $P_\Lambda$ by $\approx$20\% at considered collision energies.

The simulations are performed at fixed impact parameters, see Tab. \ref{tab:Impact}. 
To associate these impact parameters with collision centrality, one should 
keep in mind  that in the 3FD model the colliding nuclei have a shape of sharp spheres without 
the Woods-Saxon diffuse edge. This fact, implemented in the Glauber simulations 
based on the nuclear overlap calculator \cite{web-docs.gsi.de},   
results in sharp-edge centralities being $\approx 1.4$ larger than those for  
diffuse-edge case at the same impact parameter, see Tab. \ref{tab:Impact}.
\begin{table}[ht]
\begin{center}
\begin{tabular}{|c|c|c|}
\hline
b(fm) & ``sharp-edge'' centrality$^*$ & ``diffuse-edge'' centrality \\
\hline
\hline
2        & 2.5\% & 1.8\% \\
4        & 10\%  & 7\%  \\
6        & 23\%  & 16.5\%  \\
8        & 41\%  & 29\%  \\
\hline
\end{tabular}
\caption{Impact parameters of Au+Au collisions and corresponding centralities 
calculated by means of the Glauber simulations 
(the nuclear overlap calculator \cite{web-docs.gsi.de}) 
for two cases: the colliding nuclei have a 
``sharp-edge'' and ``diffuse-edge'' shape.
\\ 
$^*$ ``Sharp-edge'' nuclei are used in present simulations. 
}
\label{tab:Impact}
\end{center}
\end{table}
This way of association of impact parameters with experimental centrality
was successfully applied in previous studies of global polarization \cite{Ivanov:2020udj,Ivanov:2022ble},
light-(hyper)nuclei production 
\cite{Kozhevnikova:2023mnw,Kozhevnikova:2024itb,Kozhevnikova:2023fhh}
and directed flow \cite{Ivanov:2024gkn} in the considered energy range. 

The physical input of the present 3FD calculations
is described in Ref. \cite{Ivanov:2013wha}.
The simulations are done with two equations of
state (EoS's) with the deconfinement transitions \cite{Toneev06},
i.e. a first-order phase transition (1PT) and a crossover one. 
These EoS's give very similar predictions at  $\sqrt{s_{NN}}<$ 4 GeV  
\cite{Kozhevnikova:2023mnw,Kozhevnikova:2024itb,Kozhevnikova:2023fhh,Ivanov:2024gkn}.
However, 1PT and crossover results become differ at $\sqrt{s_{NN}}>$ 4 GeV, 
especially for the directed flow \cite{Ivanov:2024gkn}.

The thermodynamic approach to the particle polarization 
\cite{Becattini:2013fla,Becattini:2016gvu,Fang:2016vpj,Becattini:2021suc,Becattini:2021iol}
is based on the assumption of local thermal equilibrium that is not evident at moderately 
relativistic energies because the collision dynamics becomes less equilibrium with the collision energy decrease.  
Nevertheless,  it was found that 
the equilibrium is achieved at the freeze-out stage \cite{Ivanov:2022ble,Ivanov:2019gxm,Inghirami:2022afu}. 
The success of the statistical model \cite{Andronic:2005yp}
at moderate energies also indicates the thermalization at the freeze-out.

\subsection{Collision Energy Dependence}
\label{Collision Energy}

\begin{figure}[bht]
\includegraphics[width=7.4cm]{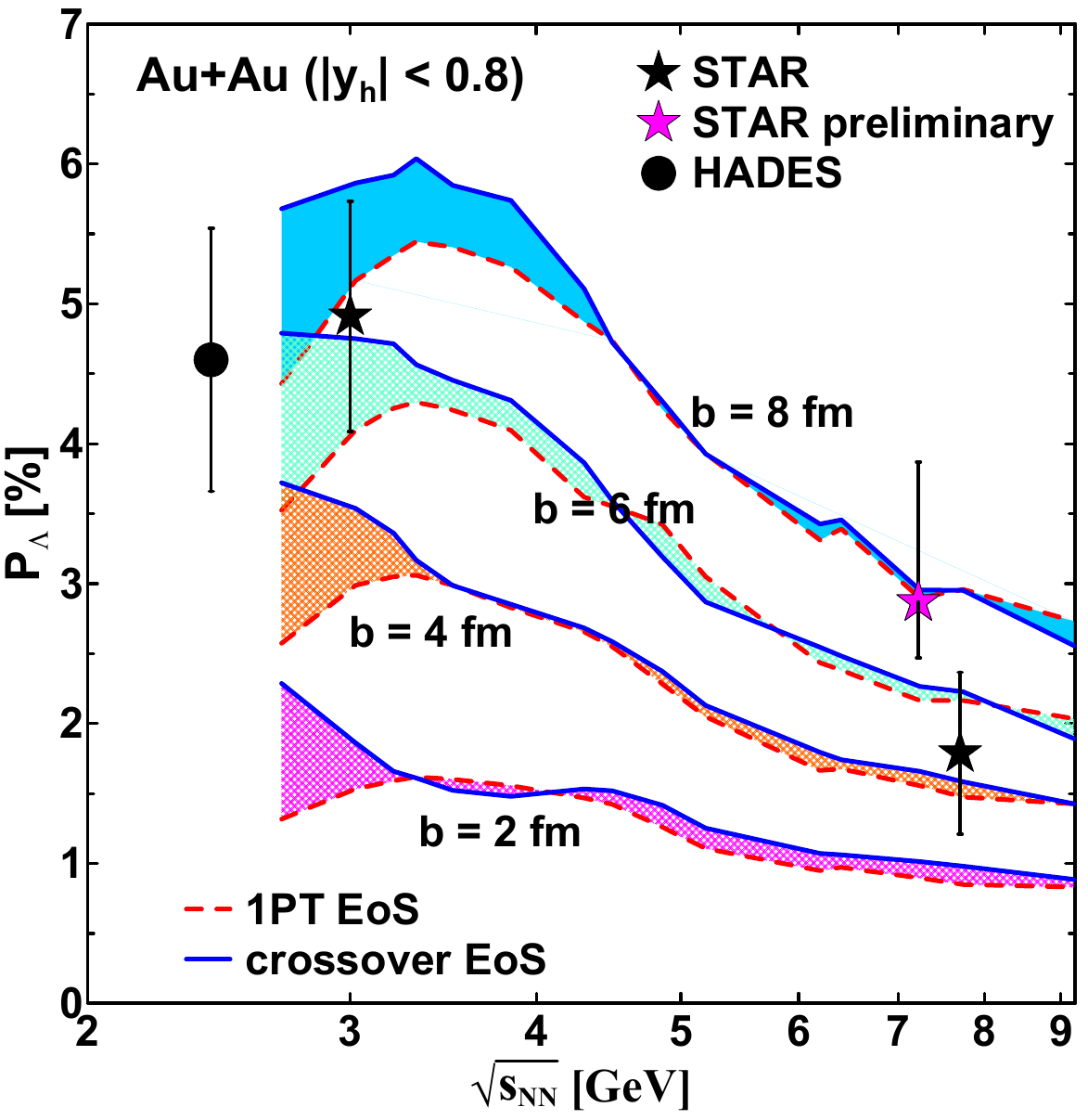}
 \caption{(Color online)
 Global $\Lambda$ polarization in midrapidity region $|y_h|<0.8$
in Au+Au collisions at different impact parameters ($b$) as 
function of collision energy $\sqrt{s_{NN}}$. 
Results for different EoS's are presented.
Areas between results for different EoS's at the same $b$ are shaded. 
Data are from Refs. \cite{STAR:2017ckg,STAR:2021beb,Okubo:2021dbt} (STAR) 
and \cite{HADES:2022enx} (HADES). 
}
\label{fig:pL-vs-sNN-b-dy8}
\end{figure}

The global $\Lambda$ polarization in midrapidity regions $|y_h|<0.8$
in Au+Au collisions at different impact parameters ($b$) as 
function of collision energy $\sqrt{s_{NN}}$ is presented in Fig. \ref{fig:pL-vs-sNN-b-dy8}.  
Available data from Refs. \cite{STAR:2017ckg,STAR:2021beb,Okubo:2021dbt} (STAR) 
and \cite{HADES:2022enx} (HADES) are also demonstrated. 
These data were obtained in measurements with different acceptances, see Tab.  \ref{tab:exp}. 
The simulations with $b=$ 6--8 fm and $|y_h|<$ 0.8 (or $|y_h|<$ 0.6 for HADES \cite{HADES:2022enx}) most closely match these acceptance conditions, with the caveat that $y_h$ is the hydrodynamic rapidity, not the true one. 
These symmetric windows are not that  good for the low-energy data 
\cite{STAR:2021beb,Okubo:2021dbt,HADES:2022enx}, 
where rapidity acceptance is asymmetric with respect to the midrapidity. However, it is still  suitable
in view of quite flat rapidity dependence of the observed $P_\Lambda$. 
\begin{table}[ht]
\begin{center}
\begin{tabular}{|c|c|c|c|}
\hline
$\sqrt{s_{NN}}$ (GeV) & exp. centrality & $b$ (fm) & (pseudo)rapidity\\
\hline
\hline
2.4 \cite{HADES:2022enx}       & 20--40\%  & 6--8  & -0.5 $<y<$ 0.3 \\
3 \cite{STAR:2021beb}          & 20--50\%  & 6--9  & -0.2 $<y<$ 1 \\
7.2 \cite{Okubo:2021dbt}       & 20--60\%  & 6--10 & -1.5 $<\eta<$ 0  \\
7.7 \cite{STAR:2017ckg}        & 20--50\%  & 6--9  &      $|\eta|<$ 1 \\
\hline
\end{tabular}
\caption{Acceptances of data available at collision energies of 2.4 $\leq\sqrt{s_{NN}}\leq$ 7.7 GeV: 
\cite{STAR:2017ckg,STAR:2021beb,Okubo:2021dbt} (STAR) and \cite{HADES:2022enx} (HADES). 
Experimental centralities are associated with impact parameters of Au+Au collisions 
by means of the Glauber simulations \cite{web-docs.gsi.de} 
for colliding ``sharp-edge'' nuclei.
}
\label{tab:exp}
\end{center}
\end{table}
%
%
%
%
%
\begin{figure}[bht]
\includegraphics[width=7.4cm]{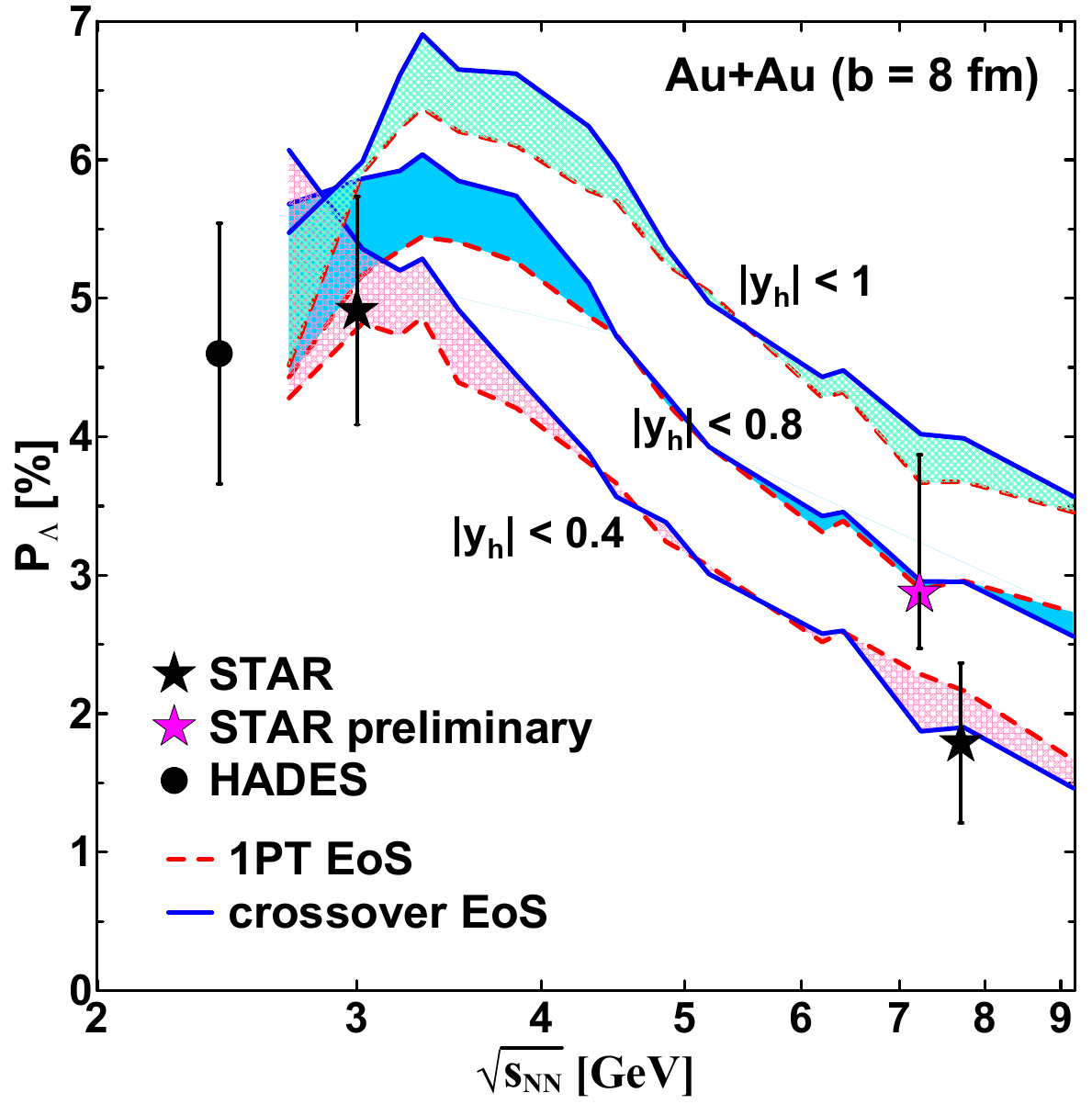}
 \caption{(Color online)
 Global $\Lambda$ polarization in different midrapidity regions $|y_h|$
in Au+Au collisions at $b=$ 8 fm as 
function of collision energy $\sqrt{s_{NN}}$. 
Results for different EoS's are presented.
Areas between results for different EoS's at the same $|y_h|$ are shaded. 
Data are from Refs. \cite{STAR:2017ckg,STAR:2021beb,Okubo:2021dbt} (STAR) 
and \cite{HADES:2022enx} (HADES). 
}
\label{fig:pL-vs-sNN-dy-b8}
\end{figure}

As seen from Fig. \ref{fig:pL-vs-sNN-b-dy8}, results that are obtained with 1PT and crossover EoS's
are very close except for the region of low collision energies $\sqrt{s_{NN}}\leq$ 3 GeV, 
where they start to diverge. At the same time all other observables, bulk and flow ones, are either 
identical or very similar within the 1PT and crossover scenarios at $\sqrt{s_{NN}}\leq$ 3 GeV
\cite{Kozhevnikova:2023mnw,Kozhevnikova:2024itb,Kozhevnikova:2023fhh,Ivanov:2024gkn,Ivanov:2013wha,Ivanov:2013yqa,Ivanov:2013yla}
because the 1PT and crossover EoS's are very similar in the corresponding range of the phase diagram. 
Therefore, the diverge of $P_\Lambda$ predicted by the 1PT and crossover scenarios
indicates enhanced numerical fluctuations in derivative computation at low collision energies. 
This is why I avoid doing $P_\Lambda$ predictions for $\sqrt{s_{NN}}<$ 3 GeV. 
Here and below, the areas between corresponding results for different EoS's are shaded in order 
to visualize the accuracy of the calculation. 

The calculation shows that a broad maximum of $P_\Lambda$ is reached at $\sqrt{s_{NN}}\approx$ 3--3.5 GeV 
at $b=$ 8 fm and probably at lower $\sqrt{s_{NN}}$ at smaller $b$. 
The $P_\Lambda$ increases with rising $b$ because of increase of the angular momentum accumulated in the 
participant matter. 

The global $\Lambda$ polarization in different midrapidity regions $|y_h|$
in Au+Au collisions at $b=$ 8 fm as function of collision energy $\sqrt{s_{NN}}$
is displayed in Fig. \ref{fig:pL-vs-sNN-dy-b8}. The maximum of $P_\Lambda$ between 3.2 and 3.9 GeV 
becomes more pronounced at $|y_h|<1$. It moves to lower energies in narrower rapidity windows.

\subsection{Rapidity dependence}
\label{Rapidity}

The fixed-target measurements \cite{STAR:2021beb,HADES:2022enx} 
also give information on the rapidity dependence of the global polarization. 
The first predictions of the rapidity dependence of $P_\Lambda$ in 
experiments within STAR-FXT program
were made in Ref. \cite{Guo:2021udq} within A Multi-Phase Transport (AMPT) 
model \cite{Lin:2004en,Lin:2014tya}. 
Although these predictions do not quantitatively agree with the STAR-FXT data 
at 3 GeV \cite{STAR:2021beb},  it is instructive to qualitatively consider them.
While the global polarization reveals quite strong increase 
from the midrapidity to forward/backward rapidities at 5 GeV $\leq\sqrt{s_{NN}}\leq$ 27 GeV,  
at $\sqrt{s_{NN}}=$ 3 GeV it turns to decrease from the midrapidity to peripheral rapidities,  
whereas the experimental dependence is quite flat at both 3 \cite{STAR:2021beb} and 7.2 GeV \cite{Okubo:2021dbt}. 
%
%
%
\begin{figure}[bht]
\includegraphics[width=7.2cm]{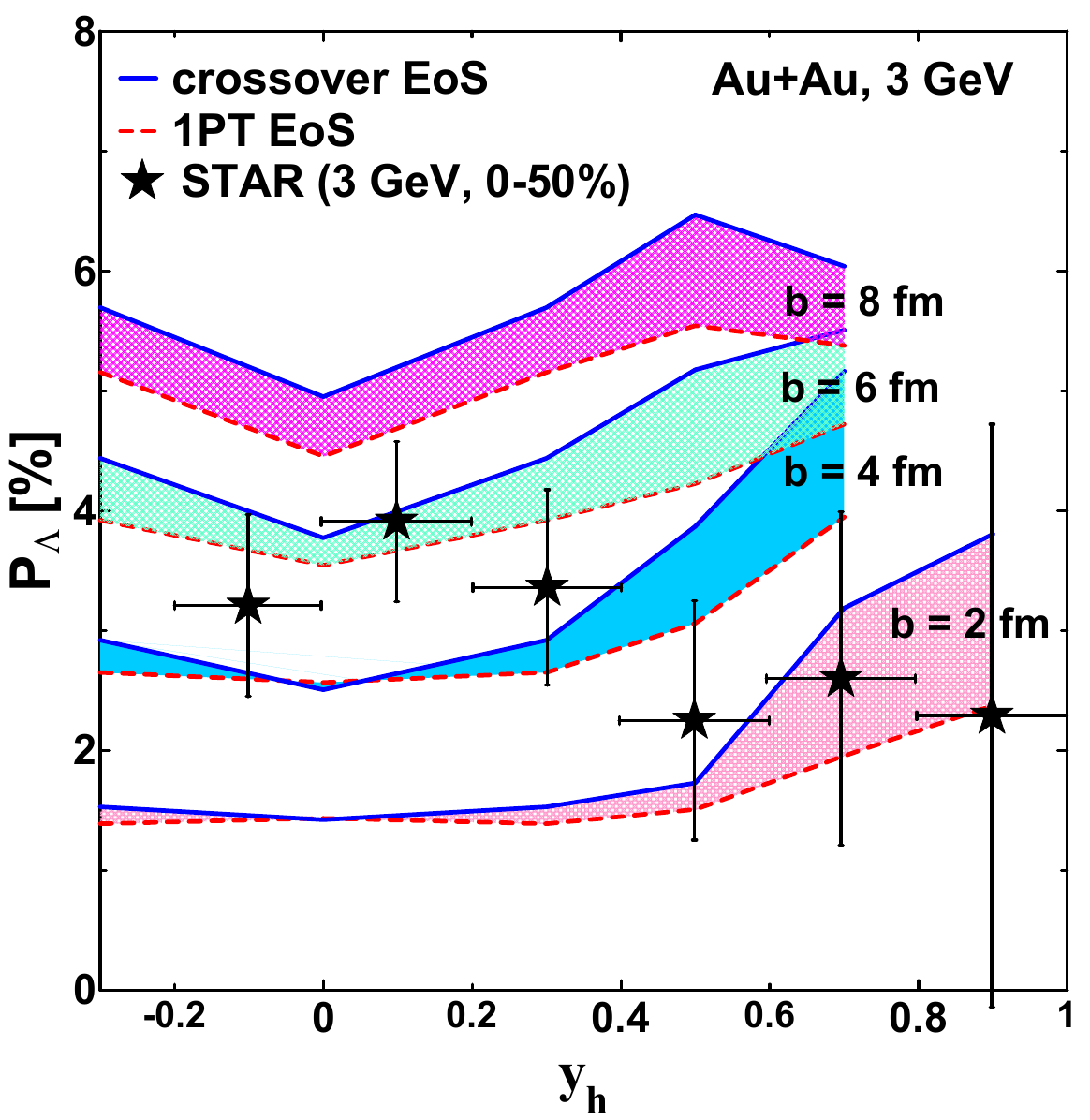}
 \caption{(Color online)
Rapidity dependence of the global $\Lambda$ polarization
in Au+Au collisions at  $\sqrt{s_{NN}}=$ 3 GeV  and different centralities 
(impact perameters $b$).
Results for different EoS's are presented.
Areas between results for different EoS's at the same $b$ are shaded. 
Data from Ref. \cite{STAR:2021beb} (STAR) correspond to centrality 0--50\%, 
i.e. $b=$ 0--9 fm. 
}
\label{fig:PL-vs-y-b-3GeV}
\end{figure}
%
%
\begin{figure}[ht]
\includegraphics[width=7.2cm]{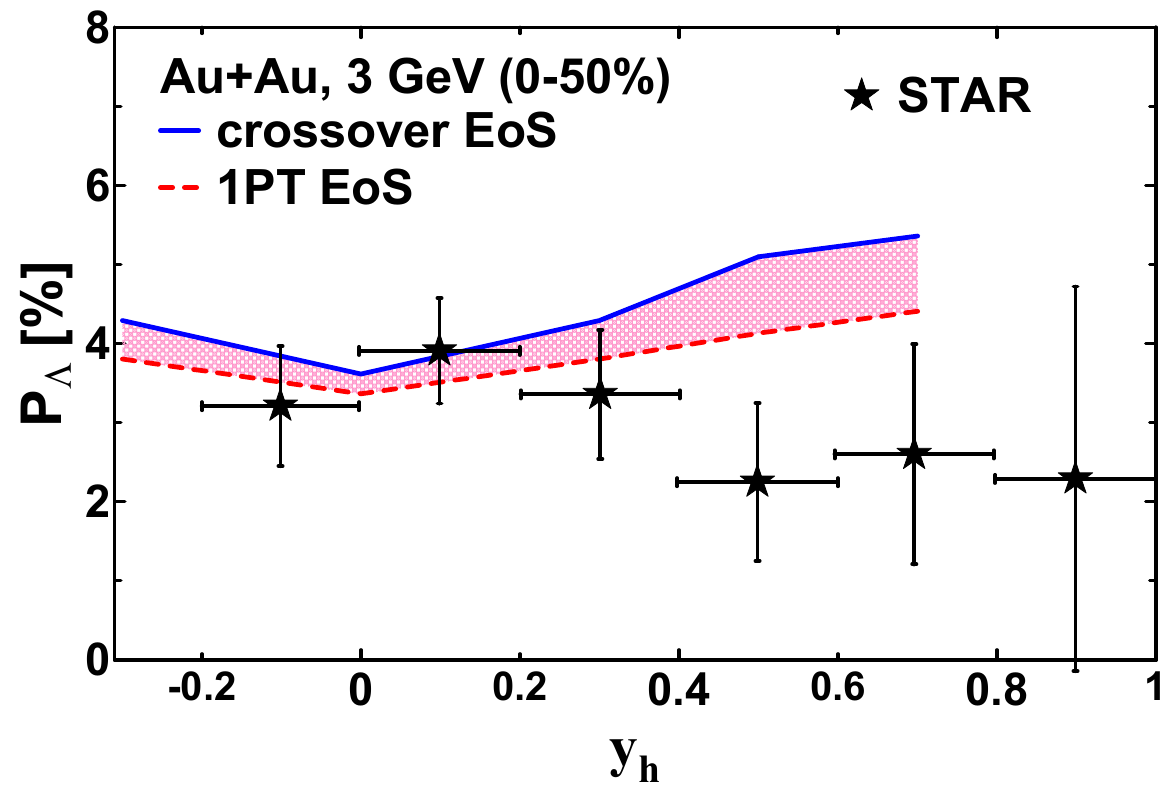}
 \caption{(Color online)
Rapidity dependence of the global $\Lambda$ polarization
in Au+Au collisions at  $\sqrt{s_{NN}}=$ 3 GeV and centrality 0-50\%. 
Results for different EoS's are presented.
Data are from Ref. \cite{STAR:2021beb} (STAR). 
}
\label{fig:PL-vs-y-3GeV_0-50pc}
\end{figure}
\begin{figure}[bht]
\includegraphics[width=7.0cm]{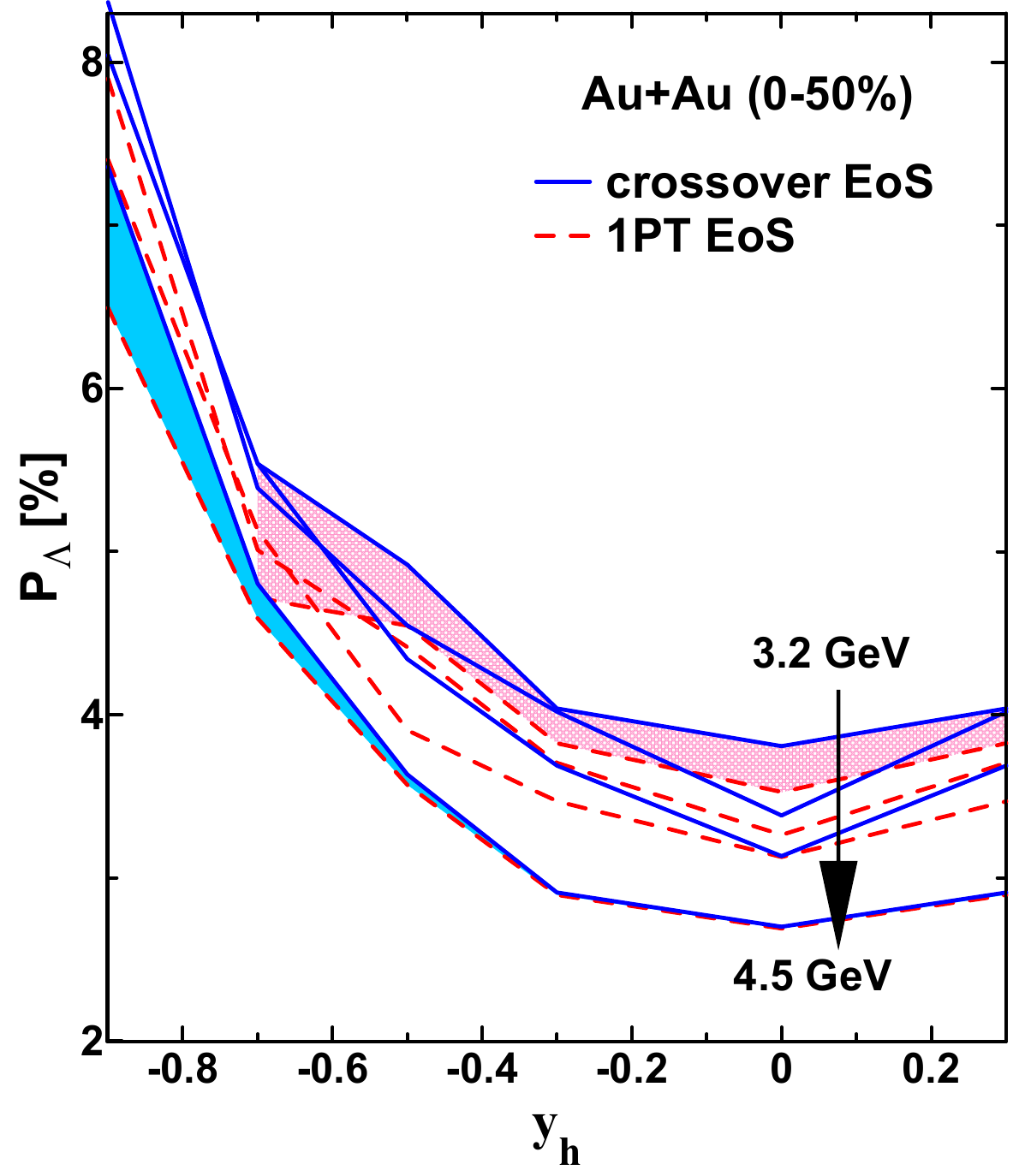}
 \caption{(Color online)
The same as in Fig. \ref{fig:PL-vs-y-3GeV_0-50pc} but 
for  $\sqrt{s_{NN}}=$ 3.2, 3.5, 3.9, and 4.5 GeV.
Areas between results for different EoS's at energies of 3.2 and 4.5 GeV are shaded. 
}
\label{fig:PL-vs-y-sNN-FTX_0-50pc-1}
\end{figure}
\begin{figure*}[bht]
\includegraphics[width=17.7cm]{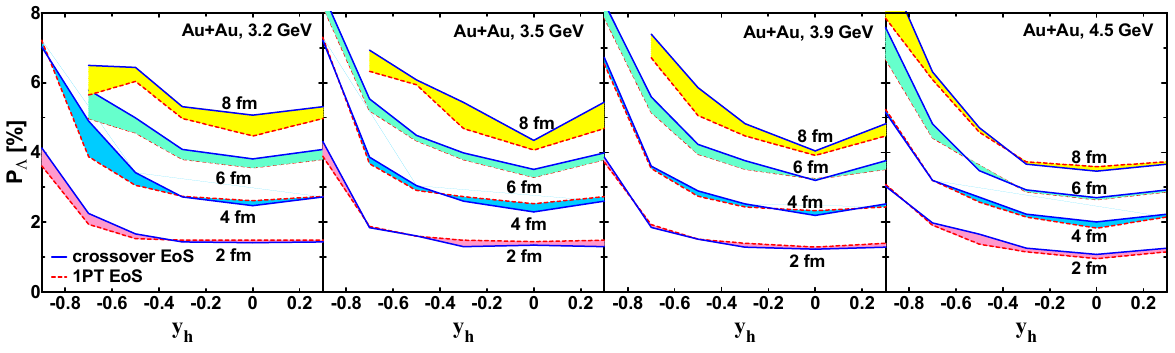}
 \caption{(Color online)
The same as in Fig. \ref{fig:PL-vs-y-b-3GeV} but 
for  $\sqrt{s_{NN}}=$ 3.2, 3.5, 3.9, and 4.5 GeV.
}
\label{fig:PL-vs-y-b-sNN}
\end{figure*}
At the same time, simulations within the 3FD model 
result in polarization increase 
from the midrapidity to forward/backward rapidities, in contrast to that in the AMPT model. 
This is a consequence of different vortex dynamics within the AMPT and 3FD models. 
In the 3FD model, the vorticity is 
mainly located at the border between participants and spectators \cite{Ivanov:2017dff},
which results in rise of $P_\Lambda$ from the midrapidity to forward/backward rapidities. 
In particular, it may result in formation of so-called vortex rings 
\cite{Ivanov:2018eej,Xia:2018tes,Lisa:2021zkj,Ivanov:2022blx,Tsegelnik:2022eoz}.

The calculated rapidity dependence of $P_\Lambda$
in Au+Au collisions at  $\sqrt{s_{NN}}=$ 3 GeV  at different impact perameters $b$
(i.e., centralities, see Tab. \ref{tab:Impact}) is presented in Fig. \ref{fig:PL-vs-y-b-3GeV}. 
Again, the shaded areas between results for different EoS's demonstrate the accuracy of the calculation. 
Results for larger impact parameters ($b\geq$ 4 fm) are displayed only up to $y_h=$ 0.7 because 
the calculation accuracy becomes poor at $y_h>$ 0.7. 
The presented STAR \cite{STAR:2021beb} data correspond to centrality of 0--50\%, 
i.e. $b=$ 0--9 fm, see table \ref{tab:exp}, which makes it difficult to draw conclusions 
about the degree of reproduction of the data because 
such a wide range of impact parameters cannot be represented by a single impact parameter. 
Therefore, one needs to perform averaging over $b$: 
\begin{eqnarray}
\label{PL-mean-def}
 \langle P_\Lambda \rangle
=  \int_0^{b_{max}} b\; d b \;P_\Lambda (b)\; / \int_0^{b_{max}} b\; d b
\end{eqnarray}
where $b_{max}=$ 9 fm.
Actual 3FD simulations of Au+Au collisions were performed at discrete 
impact parameters $b=$ 2, 4, 6, and 8 fm. Therefore, the integral in Eq. (\ref{PL-mean-def}) 
is replaced by a sum over impact parameters
\begin{eqnarray}
\label{PL-mean}
\!\!\!\!\! \langle P_\Lambda \rangle \approx
\sum_{b_i=\rm{2,4,6,8 fm}} \!\!\!\! b_i  \; P_\Lambda (b_i)\; / 
\sum_{b_i=\rm{2,4,6,8 fm}} \!\!\!\! b_i, 
\end{eqnarray}
where $\Delta b$ is canceled because $b_i$ points are equidistant.

The rapidity dependence  of the global $\Lambda$ polarization
at $\sqrt{s_{NN}}=$ 3 GeV, calculated accordingly to Eq. (\ref{PL-mean}),  is
shown in Fig. \ref{fig:PL-vs-y-3GeV_0-50pc}. This calculation well 
reproduces the  experimental rapidity dependence in the midrapidity range ($|y_h|\leq$ 0.3), while it overestimates the data
at forward rapidities.  Predictions of the rapidity dependence  of $P_\Lambda$
in Au+Au collisions at  $\sqrt{s_{NN}}=$ 3.2, 3.5, 3.9, and 4.5 GeV in centrality range of 0--50\%
are displayed in Fig. \ref{fig:PL-vs-y-sNN-FTX_0-50pc-1}. The shapes of the rapidity distributions at 3.5, 3.9, and 4.5 GeV
turn out to be very similar (up to numerical uncertainties). Only the overall normalization decreases with the collision 
energy increase. It is difficult to draw conclusions about this shape at 3.2 GeV due to high numerical uncertainties.

The predicted rapidity dependence of $P_\Lambda$
in Au+Au collisions at  $\sqrt{s_{NN}}=$ 3.2, 3.5, 3.9, and 4.5 GeV  at different impact perameters $b$
is presented in Fig. \ref{fig:PL-vs-y-b-sNN}. 
Again, results for larger impact parameters are displayed only up to $|y_h|=$ 0.7 because of 
poor accuracy at $|y_h|>$ 0.7. The $P_\Lambda$ increases in the forward/backward rapidities, 
which implies that larger gradients need to be considered. 
Numerical evaluation of these  
large gradients on the grid results in a loss of accuracy. 
Also $P_\Lambda$ (and hence the spacial gradients) increases with $b$ rise. 
As a result, numerical accuracy is noticeably worse at $b=$ 8 fm than that at 2 fm.

%
\begin{figure}[bht]
\includegraphics[width=7.0cm]{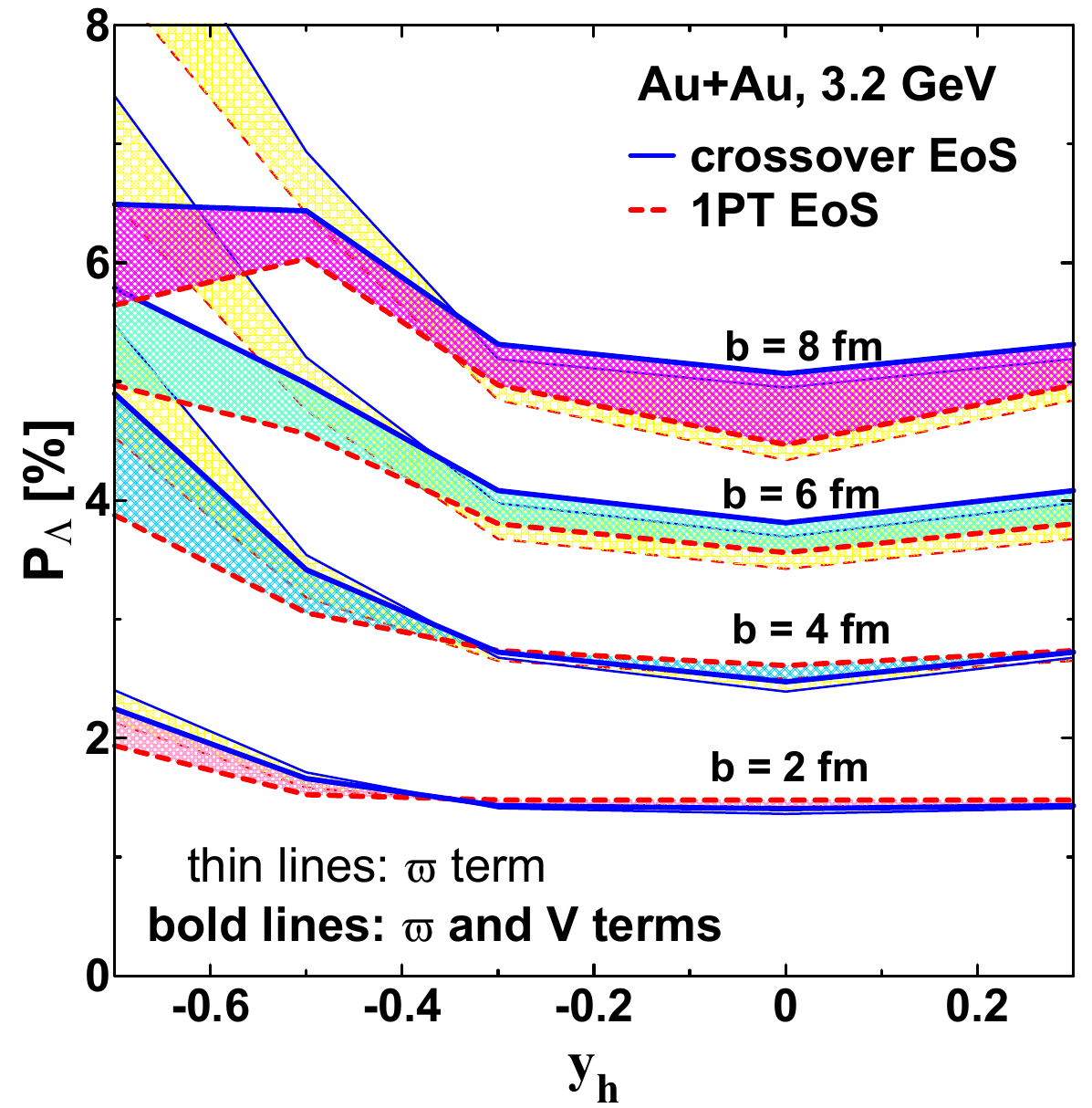}
 \caption{(Color online)
Rapidity dependence of the global $\Lambda$ polarization
in Au+Au collisions at  $\sqrt{s_{NN}}=$ 3.2 GeV  and different impact perameters $b$ 
with (bold lines) and without (thin lines) the meson-field contribution. 
Results for different EoS's are presented.
}
\label{fig:PL-vs-y-b-3,2GeV_vort-dy-FTX-V}
\end{figure}
%
%
\begin{figure}[bht]
\includegraphics[width=7.0cm]{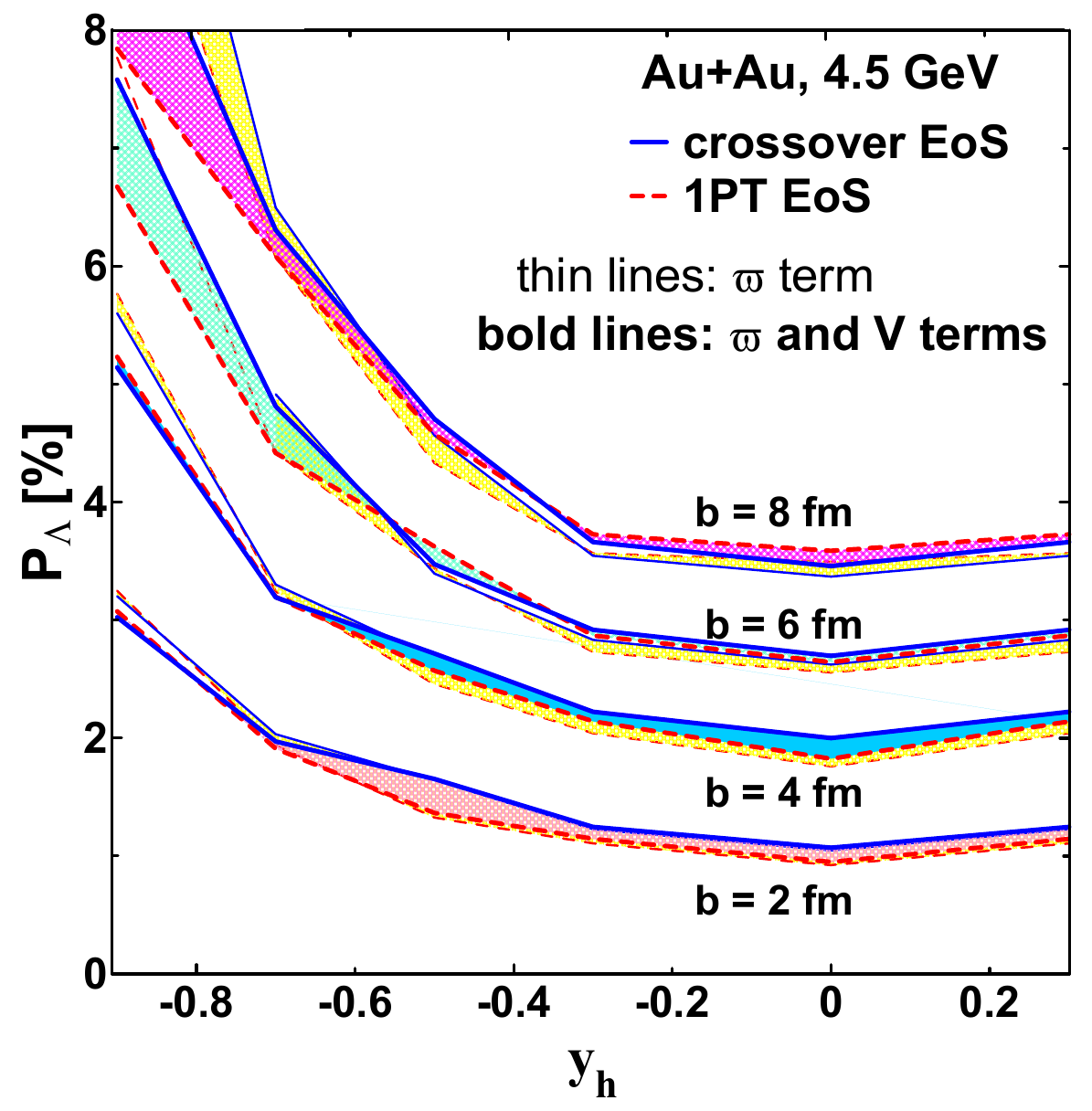}
 \caption{(Color online)
The same as in Fig. \ref{fig:PL-vs-y-b-3,2GeV_vort-dy-FTX-V} but for $\sqrt{s_{NN}}=$ 4.5 GeV.
}
\label{fig:PL-vs-y-b-4,5GeV_vort-dy-FTX-V}
\end{figure}
%

%
\begin{figure}[bht]
\includegraphics[width=7.0cm]{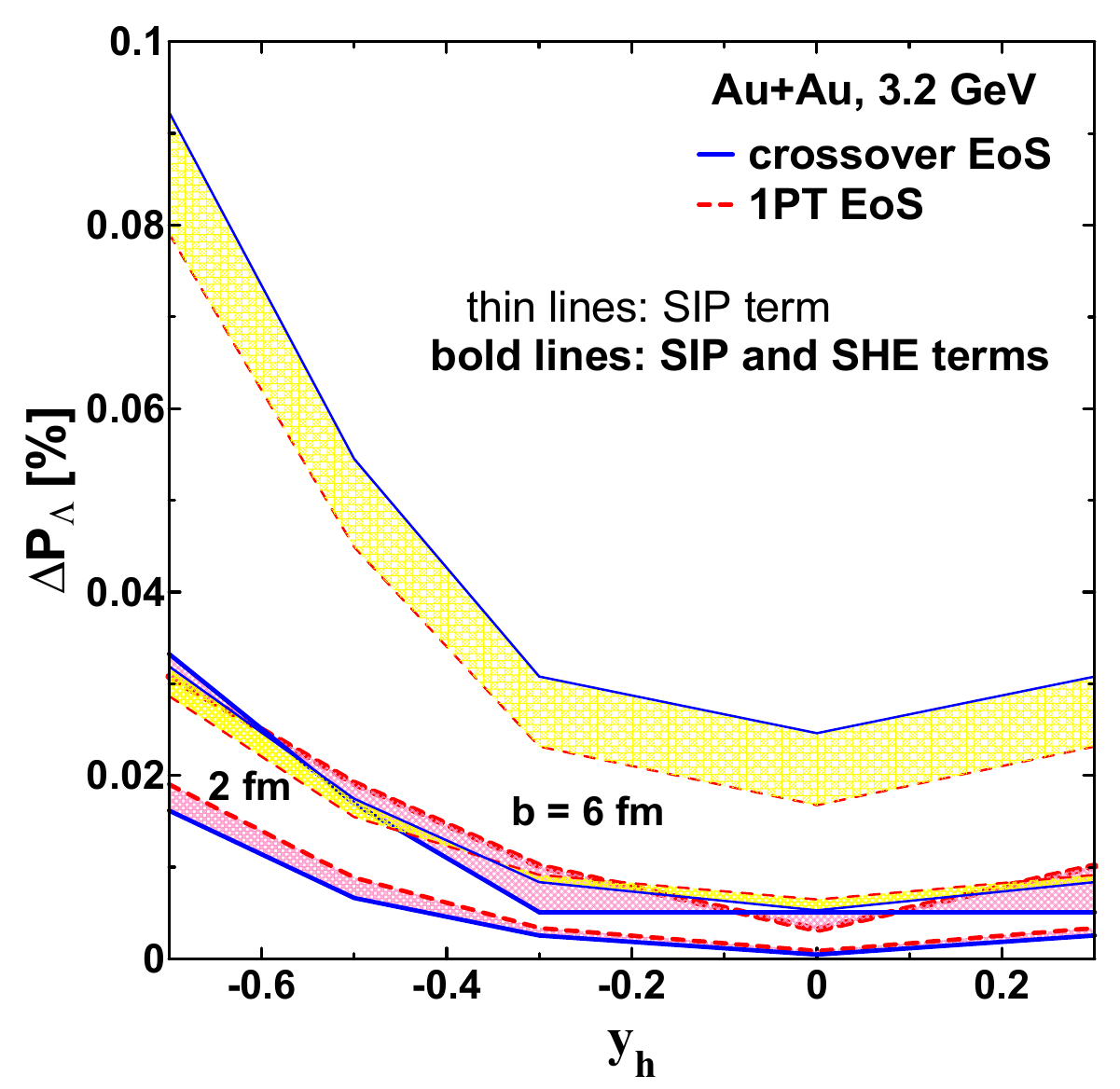}
 \caption{(Color online)
Rapidity dependence of the SIP (thin lines) and SIP+SHE (bold lines) contributions to $P_\Lambda$ 
in Au+Au collisions at  $\sqrt{s_{NN}}=$ 3.2 GeV  and different impact perameters $b$. 
Results for different EoS's are presented.
}
\label{fig:PL-vs-y-b-3,2GeV_vort-dy-FTX-XZY}
\end{figure}
%
%
\begin{figure}[bht]
\includegraphics[width=7.0cm]{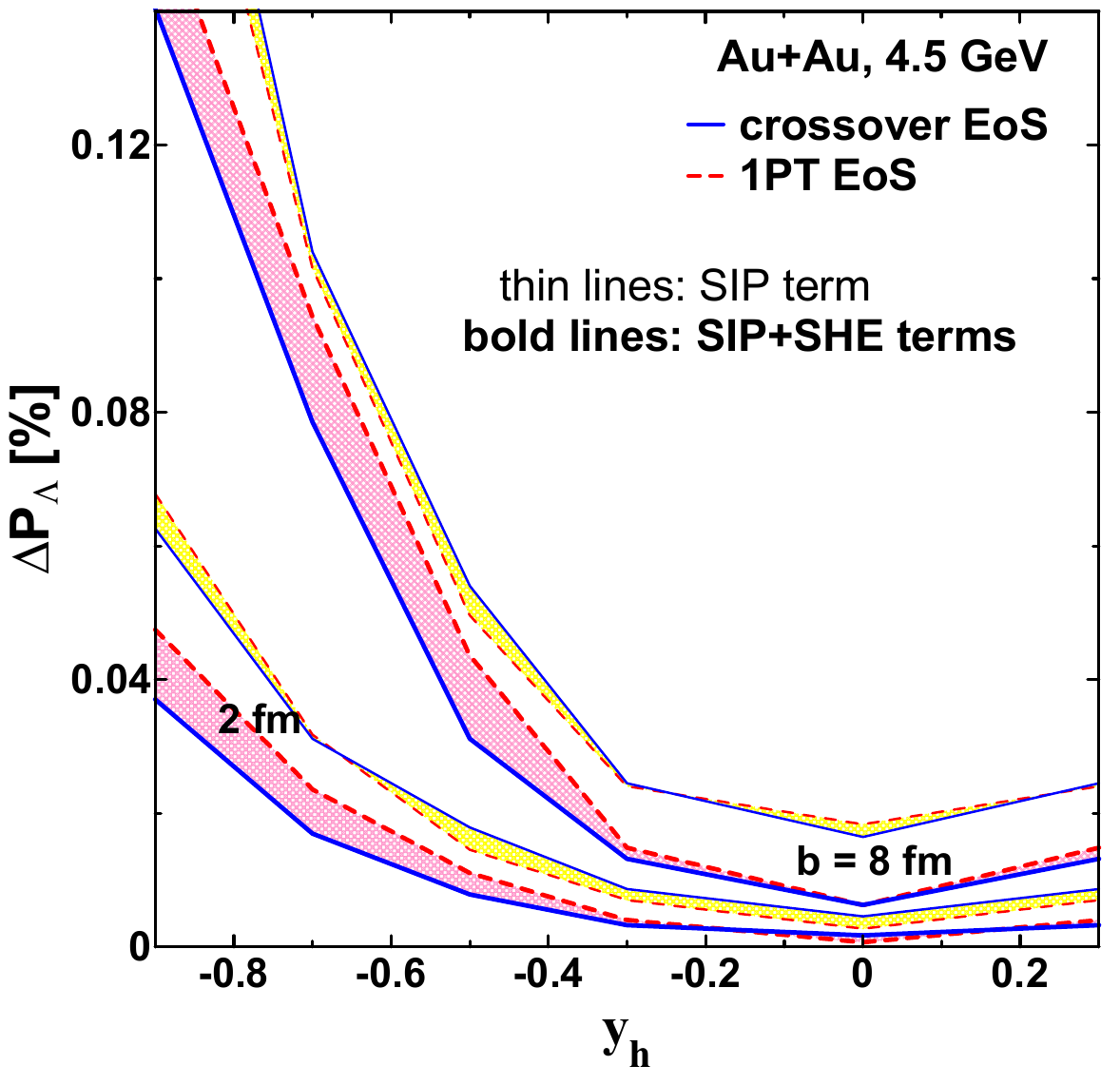}
 \caption{(Color online)
The same as in Fig. \ref{fig:PL-vs-y-b-3,2GeV_vort-dy-FTX-XZY} but for $\sqrt{s_{NN}}=$ 4.5 GeV.
}
\label{fig:PL-vs-y-b-4,5GeV_vort-dy-FTX-XZY}
\end{figure}
%

\subsection{Meson-Field Contribution}
\label{Meson-Field}


The impact of the meson-field contribution to the $P_\Lambda$, Eq. \ref{pixpgen2}, was studied in detail in Ref. 
\cite{Ivanov:2022ble} at the examples of collisions at energies of $\sqrt{s_{NN}}=$ 3 and 7.7 GeV. 
It was found that the meson-field contribution considerably flattens the 
rapidity distributions, which otherwise are strongly forward/backward peaked when only  
the thermal vorticity is taken into account.  
It improves agreement of calculated polarization with available data at 3 and 7.7 GeV. 
At the same time, the meson-field contribution is practically negligible at the midrapidity.  
Note that one of many possible parametrizations of the meson-field interaction was 
used in the present and former \cite{Ivanov:2022ble} calculations. 
It indicates the order of magnitude and character of the produced 
effect. The details may be different for other, more refined parametrizations, e.g., such as those developed in 
Refs. \cite{Maslov:2015wba,Maslov:2015msa,Hornick:2018kfi} for astrophysical applications.

Here, the effect of the meson-field contribution is demonstrated for the Au+Au collisions at energies that 
are considered in this paper, 
see Figs.  \ref{fig:PL-vs-y-b-3,2GeV_vort-dy-FTX-V} and \ref{fig:PL-vs-y-b-4,5GeV_vort-dy-FTX-V}. 
These figures demonstrate the same features as before: the meson-field contribution
considerably reduces $P_\Lambda$ at backward rapidities in semicentral collisions 
($b=$ 6 and 8 fm) at $\sqrt{s_{NN}}=$ 3.2 GeV, see Figs.  \ref{fig:PL-vs-y-b-3,2GeV_vort-dy-FTX-V}. 
The polarization remains practically unchanged in central collisions.
The influence of the meson-field contribution reduces with the $\sqrt{s_{NN}}$ rise. At 4.5 GeV, 
$P_\Lambda$ even at backward rapidities in semicentral collisions is only slightly affected by this contribution.

\subsection{SIP and SHE Contributions}
\label{SIP and SHE}

As aforementioned in sect. \ref{Local}, the SIP and SHE contributions, 
see Eqs. (\ref{P_SIP}) and (\ref{P_SHE}), have additional smallness 
$\sim T/m_{\Lambda}$ compared to that due to thermal vorticity (\ref{polint-fin}). 
Authors of Refs. \cite{Sun:2021nsg,Wu:2022mkr} found that  the SIP and SHE contributions 
to the global polarization are insignificant at the collision energies of 7.7--27 GeV.
In this subsection, the SIP and SHE effects  in the in the FXT-STAR energy range are demonstrated. 

The results of the calculations are presented in Figs 
\ref{fig:PL-vs-y-b-3,2GeV_vort-dy-FTX-XZY} and \ref{fig:PL-vs-y-b-4,5GeV_vort-dy-FTX-XZY}. 
As seen, the SIP and SHE contributions are negligible as compared with thermal-vorticity polarization. 
Moreover, the SIP and SHE contributions partially cancel each other. This calcelation is stronger at 3.2 GeV. 

%
\begin{figure}[bht]
\includegraphics[width=7.9cm]{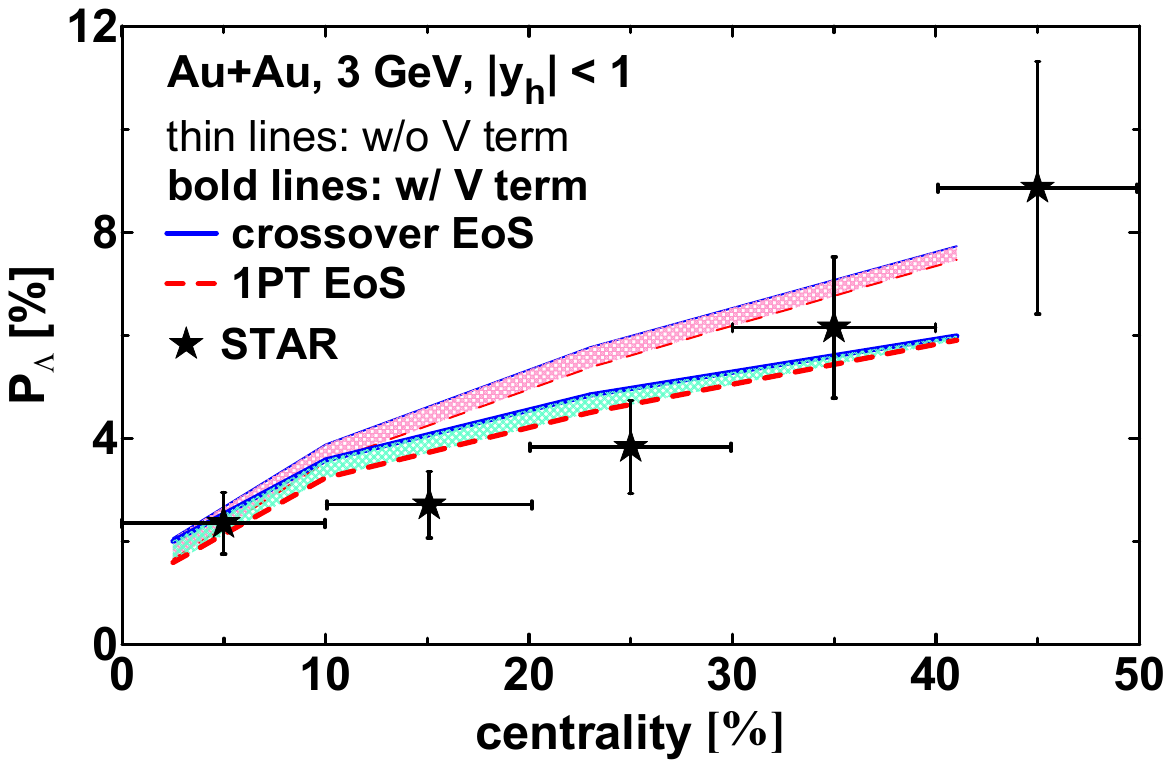}
 \caption{(Color online)
Centrality dependence of the global $\Lambda$ polarization
in Au+Au collisions at  $\sqrt{s_{NN}}=$ 3 GeV  in rapidity window $y_h\leq$ 1.
with (bold lines) and without (thin lines) the meson-field contribution ($V$)
Results for different EoS's are presented.
Areas between results for different EoS's at the same $b$ are shaded. 
Data are from Ref. \cite{STAR:2021beb}. 
}
\label{fig:PL-vs-cent_3GeV_vort-dy-FTX}
\end{figure}
\begin{figure*}[bht]
\includegraphics[width=17.7cm]{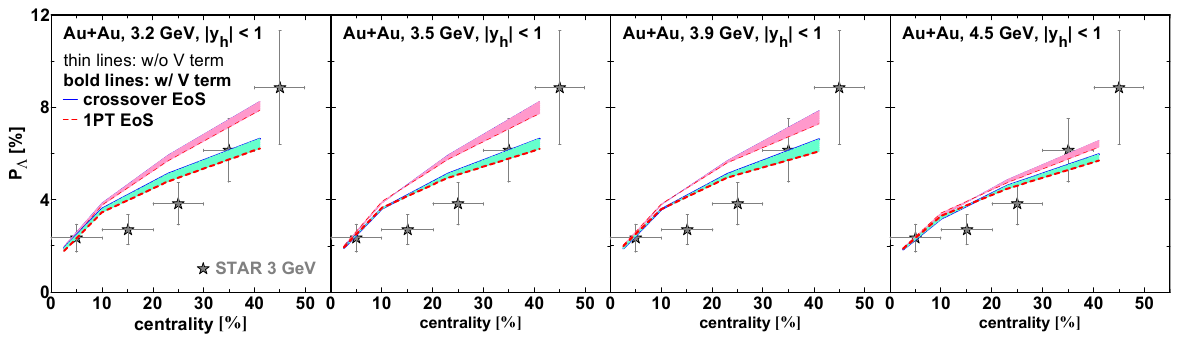}
 \caption{(Color online)
The same as in Fig. \ref{fig:PL-vs-cent_3GeV_vort-dy-FTX} but for $\sqrt{s_{NN}}=$ 3.2, 3.5, 3.9, and 4.5 GeV.
Data for $\sqrt{s_{NN}}=$ 3 GeV \cite{STAR:2021beb} are displayed to provide reference points for comparison. 
}
\label{fig:PL-vs-cent_sNN_vort-dy-FTX.pdf}
\end{figure*}
%

\subsection{Centrality Dependence}
\label{Centrality}

Centrality dependence of $P_\Lambda$ 
in Au+Au collisions at  $\sqrt{s_{NN}}=$ 3 GeV  in rapidity window $y_h\leq$ 1
is presented in Fig. \ref{fig:PL-vs-cent_3GeV_vort-dy-FTX} and compared with 
STAR data \cite{STAR:2021beb}. 
Correspondence between impact parameters and centralities is taken from Tab. \ref{tab:Impact}. 
Results with and without the meson-field contribution, Eq. \ref{pixpgen2}, are presented. 
The meson-field corrections is sizable in semicentral collisions, as it has been already pointed out 
in subsect. \ref{Meson-Field} , see Fig. \ref{fig:PL-vs-y-b-3,2GeV_vort-dy-FTX-V}. 
However, it is difficult to draw conclusions on its relevance because of large error bars of the data. 
The obtained results are in good agreement with the STAR data in very central (3--10\% centrality) and 
semi-peripheral (30--40\% centrality) collisions while overestimate the data at centralities of 10--30\%.

Predictions of the centrality dependence of $P_\Lambda$ in Au+Au collisions at  $\sqrt{s_{NN}}=$
3.2, 3.5, 3.9, and 4.5 GeV are displayed in Fig. \ref{fig:PL-vs-cent_sNN_vort-dy-FTX.pdf}. 
The STAR data \cite{STAR:2021beb} for $\sqrt{s_{NN}}=$ 3 GeV are kept in this figure as reference points  
for comparing results at different collision energies. The meson-field corrections become smaller with 
the collision energy rise. The centrality dependences (with the $V$ term)  at  3.2, 3.5, and 3.9 GeV are practically identical. 
At and 4.5 GeV, the $P_\Lambda$ the overall drop of $P_\Lambda$ starts. 
Notice that the maximum of $P_\Lambda$ is reached at 3.2--3.5 GeV 
if meson-field corrections are disregarded.

\section{Summary}
\label{Summary}

Based on the 3FD model, the global $\Lambda$  polarization 
in Au+Au collisions at moderately relativistic energies 3 $\leq\sqrt{s_{NN}}\leq$ 9 GeV 
was studied. 
Various contributions to the global  polarization were considered. 
These are contributions due to the thermal vorticity, meson field \cite{Csernai:2018yok}, 
thermal shear and spin-Hall effect \cite{Liu:2021uhn,Liu:2020dxg}. 
Feed-down from higher-lying resonances was also is taken into account, which, as found,  
reduces the polarization  by $\approx$20\%.
The results were compared with avalable data \cite{STAR:2021beb,Okubo:2021dbt,HADES:2022enx}.
These simulations closely followed the procedure described in Ref. \cite{Ivanov:2022ble}. 

Special attention was payed to the collision energies of 
$\sqrt{s_{NN}}=$ 3, 3.2, 3.5, 3.9, and 4.5 GeV, 
for which a thorough scan of the  
rapidity and centrality dependence of the global $\Lambda$  polarization was performed.
The results for 3 GeV reasonably well reproduced the corresponding STAR data \cite{STAR:2021beb}. 
While the corresponding results at $\sqrt{s_{NN}}=$ 3.2, 3.5, 3.9, and 4.5 GeV
can be considered as predictions for the STAR-FXT experimental data that are expected in the nearest future.
They are also relevant to forthcoming experiments at NICA.

The simulations were performed with two equations of
state with the deconfinement transition \cite{Toneev06},
i.e. a first-order phase transition (1PT) and a crossover one. 
Results that are obtained with 1PT and crossover EoS's
are very close except for the region of low collision energies $\sqrt{s_{NN}}\leq$ 3 GeV, 
where they start to diverge. At the same time all other observables, bulk and flow ones, are either 
identical or very similar within the 1PT and crossover scenarios at $\sqrt{s_{NN}}\leq$ 3 GeV
\cite{Kozhevnikova:2023mnw,Kozhevnikova:2024itb,Kozhevnikova:2023fhh,Ivanov:2024gkn,Ivanov:2013wha,Ivanov:2013yqa,Ivanov:2013yla}
because these  EoS's are very similar in the corresponding range of the phase diagram. 
Therefore, the diverge of $P_\Lambda$ predicted by the 1PT and crossover scenarios
indicates enhanced numerical fluctuations in derivative computation at low collision energies. 
This is why I avoided doing $P_\Lambda$ predictions for $\sqrt{s_{NN}}<$ 3 GeV. 

The global polarization increases with the collision energy decrease. 
It was predicted that 
a broad maximum of $P_\Lambda$ is reached at $\sqrt{s_{NN}}\approx$ 3--3.9 GeV 
depending on the width of the midrapidity range of observation. 
The maximum of $P_\Lambda$ between 3.2 and 3.9 GeV 
becomes more pronounced in the midrapidity window $|y|<1$. 
It moves to lower energies in narrower rapidity windows and smaller impact parameters. 
The $P_\Lambda$ increases with increasing impact parameter because of increase 
of the angular momentum accumulated in the participant matter.

It was found that  the meson-field contribution
considerably reduces $P_\Lambda$ at forward/backward rapidities in semicentral collisions 
 at 3 $\leq\sqrt{s_{NN}}\leq$ 3.9 GeV. 
The polarization remains practically unchanged in central collisions.
The influence of the meson-field contribution reduces with the $\sqrt{s_{NN}}$ rise. At 4.5 GeV, 
$P_\Lambda$ even at forward/backward rapidities in semicentral collisions is only slightly affected by this contribution. 
The results of the calculations showed that the SIP and SHE contributions are negligible as compared 
with thermal-vorticity polarization. 
Moreover, the SIP and SHE contributions partially cancel each other.

All the above results were obtained in terms of hydrodynamic rapidities, see Eq. (\ref{y}), 
rather than true rapidities of observed particles. In addition, the momentum-integrated global polarization was calculated, which does not take into account the transverse-momentum acceptance.
All this implies a semi-quantitative description of the experimental data.

In recent paper \cite{Akridge:2025jgy}, it was suggested that the global polarization 
correlates with the baryon stopping in nuclear collisions. The stronger counter-streaming baryons 
interact with each other, the more vortical collective flow they produce. The baryon stopping is indeed 
stronger at lower collision energies, which may explain the increase of the global polarization with 
decreasing collision energy. However, it is difficult to experimentally quantify the baryon stopping. 
Even at the complete stopping, the proton rapidity distribution in high-energy collisions  
has a dip in the midrapidity because of 
almost one-dimensional hydrodynamical expansion of the produced nuclear matter   
\cite{Ivanov:2012bh,Ivanov:2015vna}.  
This fact was noticed long ago by L.D. Landau in his hydrodynamical theory of multiple production of particles 
\cite{Belenkij:1955pgn}. 
This rapidity distribution is qualitatively the same as that resulting from incomplete baryon stopping.  
At the same time, the directed flow is strongly affected by the baryon stopping. 
The directed flow is a well-defined observable. 
Moreover, the correlation between the global polarization and directed flow was predicted in Refs. 
\cite{Ivanov:2020wak,Jiang:2023fad,Jiang:2023vxp,Troshin:2024nig}. 
This correlation is well measurable and may, in particular, indirectly reflect the correlation between 
the global polarization and baryon stopping.

\begin{acknowledgments} 
This work was carried out using computing resources of the federal collective usage center ``Complex for simulation and data processing for mega-science facilities'' at NRC "Kurchatov Institute" \cite{ckp.nrcki.ru}.

\end{acknowledgments}

\end{document}